# Temporal Focusing Enables Distortion-Resistant high-intensity Spatiotemporal Optical Vortices

*Jielei Ni[1], Yao Zhang[1], Qianyi Wei[1], Zhangyu Zhou[1], Shuoshuo Zhang[1,4], Yuquan Zhang[1], Qi Jin[1], Zhiyong Tan[1], Jiahui Pan[1], Ya Cheng[2,\*], Qiwen Zhan[3,\*], Xiaocong Yuan[1], and Changjun Min[1,\*]*

[1]Nanophotonics Research Center, Institute of Microscale Optoelectronics & State Key Laboratory of Radio Frequency Heterogeneous Integration, Shenzhen University, Shenzhen 518060, China

[2]The Extreme Optoelectromechanics Laboratory (XXL), School of Physics and Electronic Science, East China Normal University, Shanghai 200241, China

[3]School of Optical-Electrical and Computer Engineering, University of Shanghai for Science and Technology, Shanghai 200093, China

[3] Key Laboratory of Light Field Manipulation and Information Acquisition, Ministry of Industry and Information Technology, School of Physical Science and Technology, Northwestern Polytechnical University, Xi'an 710129, China

\* ya.cheng@siom.ac.cn, qwzhan@usst.edu.cn, cjmin@szu.edu.cn

**Abstract:**
Spatiotemporal optical vortices (STOVs) carry transverse orbital angular momentum and offer new degrees of freedom for light-matter interactions. Yet conventional focusing of STOVs introduces spatiotemporal astigmatism: the beam diffracts while the pulse duration stays constant, causing the vortex to deform away from focus. Here we overcome this limitation by introducing spectral phase modulation into a temporal focusing configuration, where angular dispersion forces the pulse to compress only at the geometric focus so that the spatial and temporal dimensions focus and defocus together. Our approach generates stable STOVs with self-similar, distortion-free evolution over an extended focal region. Besides, the orbital angular momentum vector can be continuously steered from purely longitudinal to strongly tilted orientations by adjusting the spatial dispersion, objective focal length, or input beam size. More importantly, our method offers full compatibility with high NA focusing geometry, allowing high-intensity and high-resolution applications. We validate these properties through femtosecond laser ablation under high-NA conditions and interferometric spatiotemporal field reconstruction under low-NA conditions.

1   **Introduction**

Light carries angular momentum in two fundamental forms. Spin angular momentum (SAM) comes from polarization, while orbital angular momentum (OAM) arises from a vortex like phase structure with a phase singularity [1-3]. Since the first demonstrations of optical vortices, beams with longitudinal OAM have been used to manipulate microparticles, enhance imaging, encode information in high dimensional spaces, and control interactions between light and matter from the nano to the macro scale [4-6]. More recently, these ideas have been extended from purely spatial beam shaping to fully spatiotemporal optical wave packets, in which amplitude and phase are tailored jointly in space and time [7-8]. Spatiotemporal vortices, as a basic field of singular optics in space and time, offer unique transverse OAM and novel spatiotemporal coupling, providing new control parameters for particle manipulation, ultrafast metrology, laser material processing, and potentially enabling modified light-matter interactions in strong-field physics [9–13].

Realizing this potential requires accurate control of such vortices in a focused geometry, where most high-intensity and high-resolution applications take place. The standard approach encodes a spiral phase in the frequency domain using a *4f* pulse shaper and then focuses the beam with a simple lens [10-12]. This method produces a well-defined STOV only at the exact focal plane. Away from focus, spatial and temporal evolutions become decoupled: the beam diffracts while the pulse duration remains nearly constant. This spatiotemporal astigmatism causes the vortex ring to split into two lobes and the OAM density to redistribute along the propagation axis [10]. For interactions that extend over a finite propagation length, such as high harmonic generation in gases or solids, plasma wake-field excitation, nonlinear frequency conversion in micro and nanostructures, or imaging through thick samples, this defocus-induced distortion reduces mode purity, weakens transfer of OAM [14-16].

Recent efforts have addressed propagation stability using diffraction-free beam configurations [17-19]. These approaches employ various non-diffracting spatial profiles such as Airy or Bessel structures to constrain the spatial diffraction and match it with the temporal dynamics. By distributing the optical energy along extended propagation paths rather than concentrating it at a focal point, these methods achieve remarkable stability over millimeter-to-meter distances, making them well suited to long-reach transmission and communication scenarios. However, this comes with a deliberate trade-off in tight-focusing capability and peak intensity, which makes such beams generally less tailored to strong-field interactions and high-efficiency nonlinear optics.

Here we propose and demonstrate a fundamentally different strategy: modulating temporal dynamics to match the spatial diffraction, thereby enabling simultaneous spatiotemporal focusing of STOV that binds the spatial and temporal dimensions together throughout focusing and propagation. Our approach introduces temporal focusing by angular dispersion via a grating pair, applies a spiral phase to the spatially dispersed spectrum, and recombines all frequency components at a common focal plane. Although temporal focusing has been used for nonlinear microscopy and precision micromachining [20-23], prior work only exploited its spatiotemporal confinement purely through amplitude engineering. The structured-light potential of SSTF has remained unexplored.

By fully exploring the spectral phase as a design degree of freedom, we eliminate the spatiotemporal astigmatism inherent in conventional schemes and offers three key advantages simultaneously. First, temporal focusing ensures that the STOV undergoes self-similar, distortion-free evolution throughout the focal region: both the ring-shaped intensity profile and the helical phase singularity are faithfully preserved over an extended propagation range even under a relatively high NA of 0.5, a level of propagation robustness that is difficult to achieve with conventional 4f pulse shapers. Second, the angular dispersion introduces a controllable pulse-front tilt (PFT) that enables continuous OAM vector steering, allowing the OAM vector to be rotated from purely longitudinal to strongly transverse orientations in space-time coordinates by adjusting the spatial dispersion, focal length, or input beam size. This controllability connects to recent efforts on generating spatiotemporal vortices with arbitrarily oriented OAM using astigmatic mode converters [24], spatial chirp [25], engineered photonic nodal lines [26], or the superposition of spatial and spatiotemporal vortices [27]. A distinguishing feature of our method, however, is that both the ring-like intensity distribution and the phase singularity are maintained in a self-similar form even away from the exact focus, a consequence of the simultaneously spatiotemporal scaling inherent to SSTF. Third, and crucially, unlike non-diffracting approaches, our approach preserves full compatibility with high-NA spatial focusing, so the temporal and spatial compression act in concert to yield peak intensities substantially higher than those attainable with conventional 4$f$ systems operating at comparable bandwidths. This intensity advantage, combined with the propagation stability and the continuous OAM orientational control, establishes a versatile platform for downstream applications ranging from ultrafast laser processing to enhanced light–matter interactions that demand precise command over both the direction and topology of angular momentum in tightly confined spatiotemporal volumes.

## 2  Result
### 2.1  Concept

As schematically shown in **Fig. 1(a)**, a conventional 4$f$ pulse-shaping system uses a grating, a cylindrical lens (CL) and a spatial light modulator (SLM) to imprint a helical phase in the $k_y-\Omega$ domain, with all components separated by the focal length $f$ of the cylindrical lens. After passing through the objective lens, this produces a spatiotemporal vortex in the y−t plane at the focus (4f-STOV (**Fig. 1(a), right pannel**). However, this vortex only exists in the y−t plane at focus and cannot be maintained during propagation. As the observation plane moves away from the focus, the ring-shaped intensity pattern breaks apart.

To create a more robust spatiotemporal optical vortex, we introduce spectral phase modulation into a temporal focusing configuration as shown in **Fig. 1(b)**. We disperse the spectrum spatially using a grating pair, apply a designed phase pattern across the dispersed beam, and then refocus it with an objective lens. We term the resulting structure a temporal-focusing spatiotemporal optical vortex (TF-STOV). Our approach relies on two key features. First, the angular dispersion from the gratings introduces a pulse-front tilt at the focus, which, together with the spatial vortex component inherent in the focused field, orients the vortex obliquely in three-dimensional space-time domain. The OAM now points in a tilted direction with both longitudinal and transverse components. Second, the temporal focusing mechanism ensures that the pulse duration is minimized at the spatial focus and progressively broadens away from it, thereby coupling the spatial and temporal degrees of freedom along the propagation axis.

As shown in the right panel of **Fig. 1(b)**, when scanning along the z-axis through the focal region, the TF-STOV exhibits a tilted toroidal intensity profile that preserves its ring-shaped structure at each z position. The projections onto both the x–y and y–t planes consistently display well-defined annular patterns, with the vortex cores remaining aligned throughout. This propagation robustness is remarkably different from that of 4f-STOV (**Fig. 1(a)**), where the ring-shaped intensity profile at focus rapidly distorted into two lobes upon defocusing, owing to the decoupling of spatial and temporal evolution. The analytical framework developed in Methods (Eq. (13)–(21)) formalizes this picture: the focal field naturally decomposes into co-propagating spatial and spatiotemporal vortex components (Eq. (13)), whose interference produces the tilted vortex core (Eq. (20)–(21)); the temporal focusing mechanism couples the pulse duration to the propagation coordinate (Eq. (16)–(18)), and when the temporal and spatial Rayleigh ranges are matched (Eq. (19)), the ring-shaped structure scales self-similarly throughout the focal region.

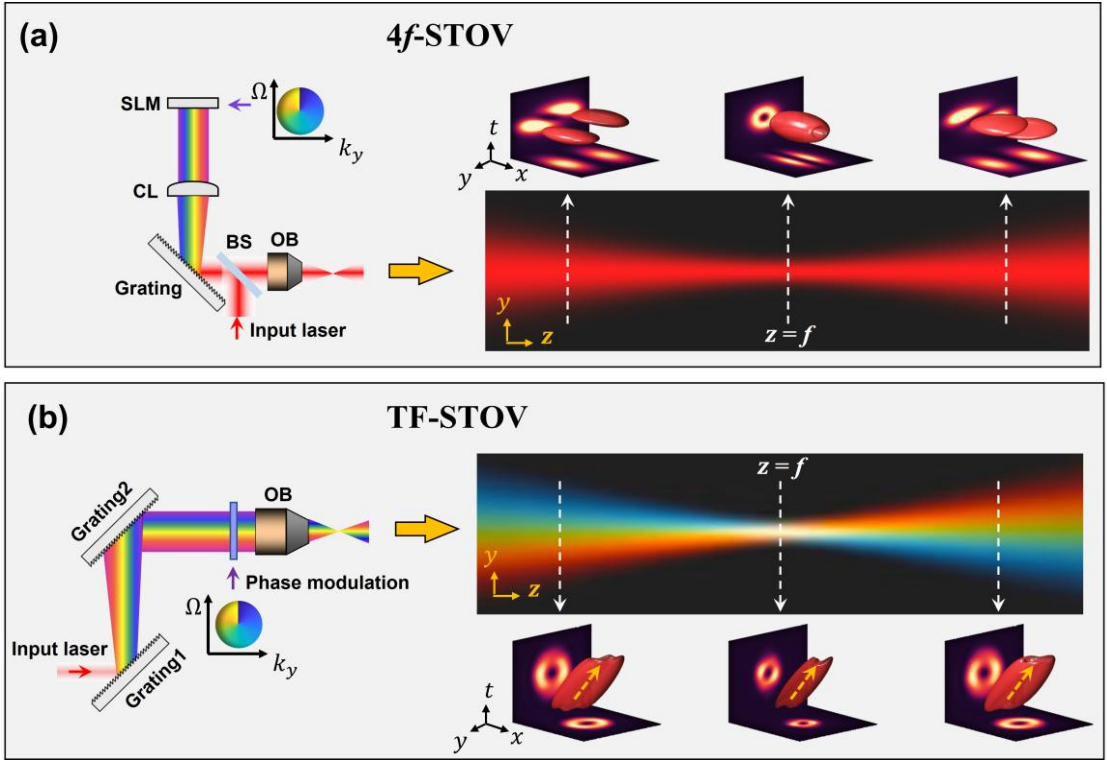

**Figure 1. Concept.** (a) A conventional 4$f$ pulse-shaping system generates a spatiotemporal vortex by imprinting a helical phase in the $k_y$–$\Omega$ domain via a spatial light modulator (SLM). The right panel shows the focused field in the $y$–$z$ plane (background), with dashed lines marking three axial positions; the corresponding $x$-$y$-$t$ spatiotemporal intensity distributions are displayed above. At focus the STOV exhibits a ring-shaped profile in the $y$–$t$ plane, but the structure rapidly distorts into two separated lobes upon defocusing. (b) TF-STOV generation by introducing spectral phase modulation into a temporal focusing configuration using a grating pair. The right panel shows the $y$–$z$ propagation field (background). The $x$-$y$-$t$ distributions at three axial positions (below) display tilted toroidal intensity profiles whose ring-shaped structure and vortex cores are well preserved throughout the focal region. Yellow arrows indicate the tilted OAM direction carrying both longitudinal and transverse components. The 2D intensity maps in the $x$-$t$ and $y$-$t$ planes represent cross-sectional slices through the 3D spatiotemporal field at $y = 0$ and $x = 0$, respectively.

## 2.2 Generation of TF-STOV

Temporal focusing is a widely used approach for achieving localized pulse compression and high peak intensity in the focal region. The key idea is to introduce angular dispersion into the spectrum, so that different frequency components travel along different directions in space and only recombine temporally at the focal plane. After dispersing the spectrum in the spatial domain, we apply a spiral phase to the frequency components using either a phase plate or a spatial light modulator (SLM), as illustrated by **Fig. 2a**. This approach effectively encodes OAM into the temporal spectrum. Subsequently, the dispersed and phase-modulated spectrum is recombined and focused by an objective lens.

Using Fresnel diffraction theory (see **Methods**), we simulated STOV generation based on the SSTF configuration. In this calculation, we employed a laser beam with a central wavelength of 800 nm and a spectral bandwidth of 10 nm (FWHM), corresponding to a transform-limited pulse width of 109 fs. The spatial chirp is set to $\alpha = 7 \times 10^{-16} m/Hz$. The incident beam waist in x- and y-direction are set to $W_x = 1\ mm$ and $W_y = 2.8\ mm$. The objective lens had a focal length of 8 mm and a numerical aperture (NA) of 0.5.

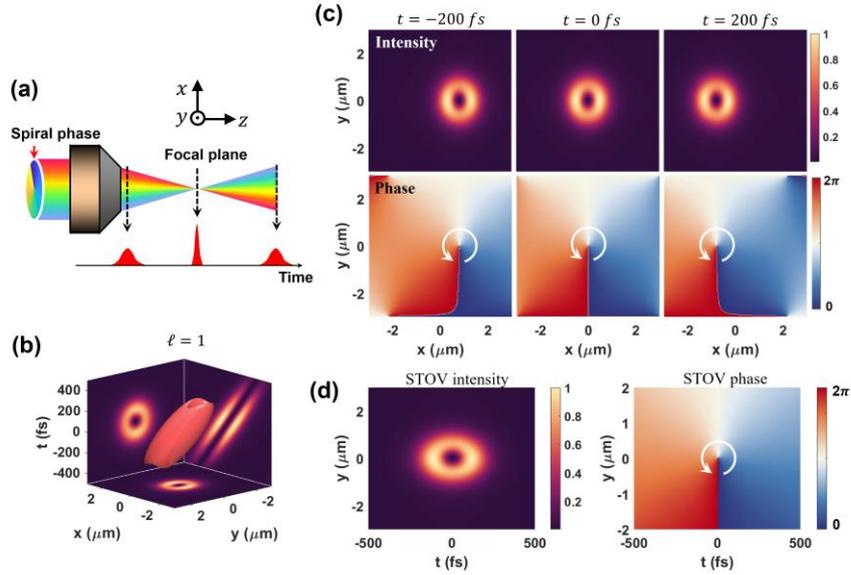

**Figure 2. Spatiotemporal optical vortex (STOV, ℓ = 1) generated of at the focal plane via SSTF.** (a) Optical schematic for SSTF. (b) At the focal plane (z = 0), simulated 3D iso-intensity surface plot shows that the generated STOV is oriented obliquely in 3D space-time domain, with ring-shaped intensity distributions in both the spatiotemporal plane (*y-t*) and the spatial plane (*x-y*). (c) Spatial intensity and phase distributions in the *x-y* plane at different time slices, demonstrating that the ring profile and the central phase singularity are well preserved across the pulse duration. (d) Intensity and phase distributions in the *y-t* plane (*x* =0) showing the characteristic spatiotemporal vortex structure with a helical phase around the intensity null.

When a spiral phase with topological charge $\ell = 1$ is applied onto the beam before the objective lens, the focal-field distribution exhibits distinct features in both spatial and temporal domains, as shown in **Fig. 2b**. The iso-intensity surface reveals a tilted hollow structure in three-dimensional *x-y-t* coordinate system, forming a tilted STOV. Cross-sections through this structure show ring-shaped intensity patterns in both the spatiotemporal plane (*y-t* plane at *x* = 0) and the spatial plane (*x-y* plane at *t* = 0). In the *x-t* plane at *y* = 0, the intensity distribution appears oblique due to the pulse-front tilt effect [28]. Examination of the transverse spatial profiles at different time slices reveals that the vortex structure remains well preserved throughout the pulse duration, maintaining both the annular intensity distribution and the helical phase pattern with a central singularity (**Fig. 2c**). This temporal consistency demonstrates that the spatial vortex topology is robust across the entire pulse envelope. In the spatiotemporal plane (*y-t* plane) at *x* = 0, the intensity and phase distributions display a typical STOV with a donut-shaped intensity profile and spiral phase, confirming the spatiotemporal vortex nature (**Fig. 2d**). Generation of higher-order STOVs are presented in Supplementary **Fig. S1**.

### 2.3 Off-focus spatiotemporal co-scaling of TF-STOV

Away from the focal point, the spatial and temporal dimensions of the STOV scale simultaneously in this work owing to the advantage of temporal focusing. **Fig. 3(a)** displays the intensity profiles of TF-STOV on the *y-t* plane (at *x* = 0) at different axial positions. Distances along z are normalized to the Rayleigh range $z_{Ry} = \frac{\lambda f^2}{\pi W_y^2}$. To present the intensity variation across 3D space, all profiles are normalized to the peak value at *z* = 0. Individually normalized views that emphasize vortex morphology at each plane are shown in **Fig. 3(e)**.

Close to focus ($z = 0$) the pulse remains tightly confined in both y and t axis, producing a compact, doughnut-shaped intensity distribution. As |z| increases the beam expands diffractively in y-axis, while similarly, the pulse broadens in t-axis because the angular dispersion. Analytically the effective spectral bandwidth narrows away from focus (Eq. (14) in Methods), causing the pulse duration near the focal region to scale as $\tau(z) \approx \tau_0\sqrt{1 + z^2/z_{Rt}^2}$ (Eq. (18) in Methods). Consequently, the STOV retains a relatively well-defined annular structure with both temporal and spatial dimensions scaling concurrently. Even at the extremes of the measurement range (|z| = $4z_{Ry}$), where the peak intensity has decreased by more than an order of magnitude, the topological vortex structure remains remarkably well preserved. This behavior is fundamentally distinct from that of STOVs generated using a 4*f* system.

**Fig. 3(b)** shows the intensity distribution on the y-t plane (at x = 0) for an STOV produced via a traditional 4*f* system. In this case, focusing and defocusing are observed only in the spatial domain. In the temporal dimension, the pulse width remains constant when focusing or defocusing. Due to the purely spatial focusing process in the 4*f* system, the annular intensity structure cannot be maintained. As the axial position moves away from the focus, the ring-shaped intensity structure gradually splits into two lobes and separates spatially.

The variation in pulse width at different axial positions is further analyzed numerically in **Fig. 3(c)** for both schemes, demonstrating temporal compression to a minimum at the focal plane (z=0) TF-STOV while the 4*f*-STOV exhibits no temporal focusing but a constant pulse duration throughout propagation. Near the focal region, the numerically extracted pulse widths of TF-STOV (red circles) agree well with the analytical prediction $\tau(z) \approx \tau_0\sqrt{1 + z^2/z_{Rt}^2}$ (blue dashed curve; Eq. (18) in Methods), confirming the validity of the analytical model for TF-STOV.

In order to quantifies how well the STOV structure is preserved during propagation, we performed intensity correlation analysis (see **Methods**). After normalizing each distribution and rescaling it by the pulse duration ratio, we calculate the 2D Pearson correlation coefficients between spatiotemporal intensity distributions at different *z*-positions and the reference distribution at *z* = 0. The results are shown in **Fig. 3(d)**. Under a relatively high NA of 0.5, the intensity correlation of the TF-STOV remains above 0.9 over a propagation range of $-3.25z_{Ry} \leq z \leq 3.25z_{Ry}$ (a total extent of $6.5z_{Ry}$), compared with $-z_{Ry} \leq z \leq z_{Ry}$ ($2z_{Ry}$) for the 4*f*-STOV, representing a 3.25-fold enhancement in the effective propagation distance over which the vortex structure is preserved. The higher correlation maintained in the TF-STOV configuration reveals that temporal focusing preserves the STOV structure more effectively during propagation. Analytically, optimal self-similar co-scaling requires equal temporal and spatial Rayleigh ranges (Eq. (19) in **Methods**). The simulation parameters used here approximately satisfy this matching condition, which explains the high structural correlation observed over the extended focal region.

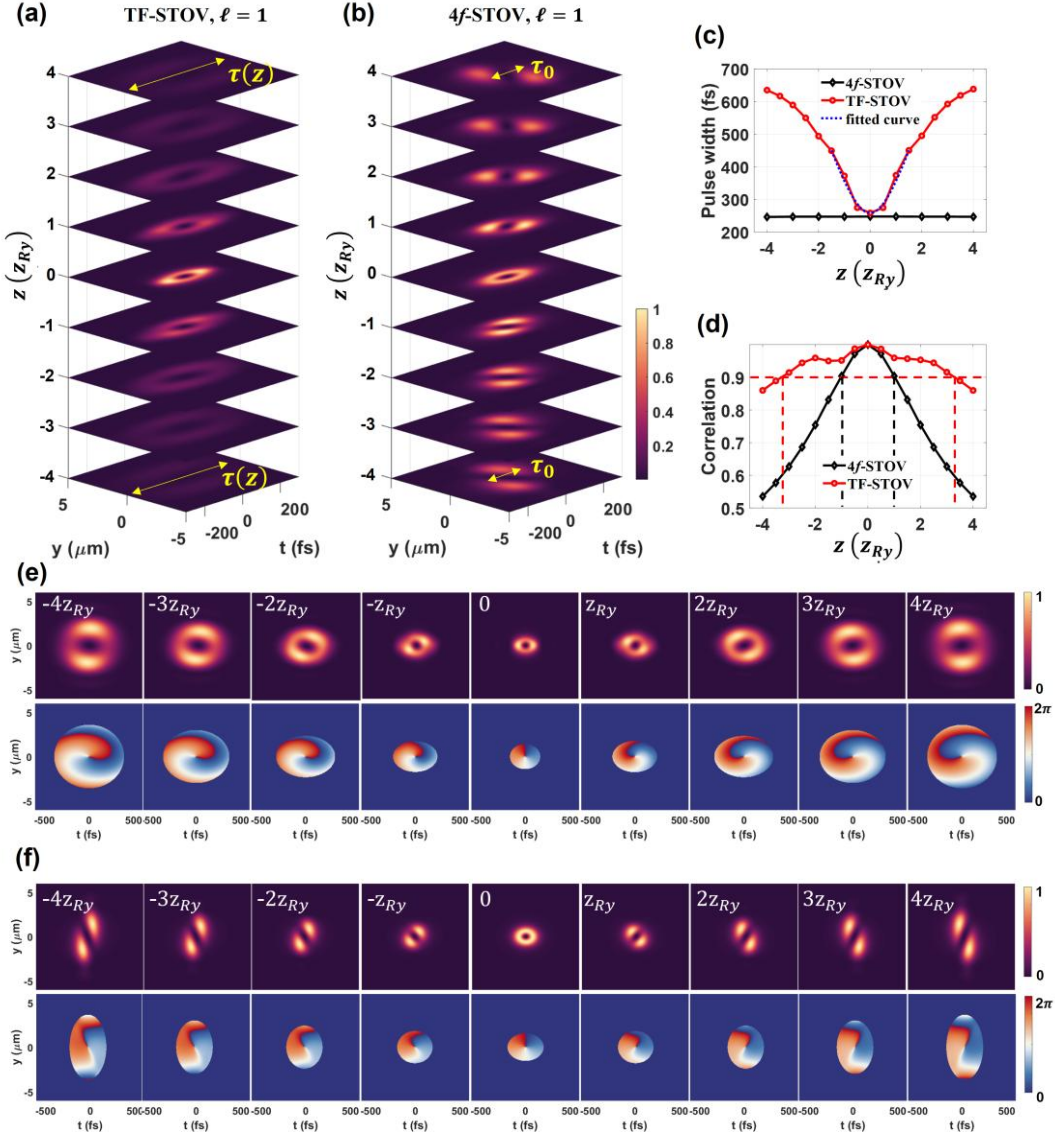

**Figure 3. Axial evolution comparison between TF-STOV and 4*f*-STOV.** (a) Intensity distributions in the spatiotemporal plane (*y-t*) at different propagation distances (normalized to Rayleigh range $z_{Ry}$) for TF-STOV, showing focusing and defocusing in both spatial and temporal dimensions. (b) Intensity distributions in the *y-t* plane at different $z/z_{Ry}$ for 4*f*-STOV, where the temporal dimension remains unfocused and the intensity pattern rapidly splits into two lobes as the spatial dimension defocuses. (c) Pulse duration as a function of propagation distance in both systems, demonstrating temporal compression to a minimum at the focal plane ($z = 0$) for TF-STOV while 4*f*-STOV exhibits no temporal focusing with constant pulse duration throughout propagation. (d) Intensity correlation coefficients between spatiotemporal distributions at different *z*-positions and the reference distribution at $z = 0$, quantifying the self-similarity of propagation in both systems. The higher correlation maintained in TF-STOV reveals that temporal focusing preserves the STOV structure more effectively during propagation. (e-f) Individually normalized intensity and phase evolution of TF-STOVs ($\ell = 1$) (e) and 4*f*-STOVs (f) during propagation. Each panel here is individually normalized to its local maximum intensity to emphasize the vortex morphology and phase structure at each propagation distance.

The propagation characteristics of second-order STOVs are presented in **Supplementary Fig. S2**. The results confirm that the self-similar propagation behavior extends to higher-order STOVs. In particular, upon defocusing, the second-order STOV generated by the 4$f$ scheme undergoes pronounced singularity splitting, with the single charge-2 vortex breaking into two spatially separated unit-charge vortices. In contrast, the TF-STOV exhibits significantly suppressed splitting, maintaining a more intact higher-order vortex structure over a comparable propagation range.

### 2.4 Arbitrary OAM orientation of TF-STOV

In our scheme, the OAM orientation of the TF-STOV at the focal plane can be precisely controlled by tuning three key system parameters: the spatial dispersion factor $\alpha$, the focal length of the objective $f$, and the incident beam waist $W_x$. The orientation is quantified by the slope $k_t$ of the singularity line in the x-t plane ($k_t = \Delta t / \Delta x$), with the singularity trajectory indicated by white traces in each panel of **Fig. 4**. Notably, $k_t = 0$ corresponds to a transverse STOV whose OAM is purely oriented along the propagation direction, whereas $k_t \to \infty$ recovers a conventional spatial optical vortex with OAM aligned along the optical axis. By continuously varying the above parameters, the singularity line can be smoothly rotated between these two limiting cases, enabling full control over the OAM orientation in the spatiotemporal domain.

**Figs. 4(a1-a4)** show the dependence on the spatial dispersion factor $\alpha$. Three representative cases with different dispersion levels are visualized in **Figs. 4(a1-a3)**, while **Fig. 4(a4)** provides quantitative analysis of the slope ($k_t$) versus spatial dispersion factor ($\alpha$). Notably, the TF-STOV orientation exhibits a non-monotonic relationship with spatial dispersion, distinct from the linear scaling of pulse-front tilt reported by He et al. [28] (Red line, **Fig. 4(a4)**). This behavior is quantitatively captured by the analytical expression for singularity slop (Eq. (21) in **Methods**), which reveals a competition between a $1/\alpha$ term arising from spatial modulation and a linear $\alpha$ term from pulse front tilting. Without spatial dispersion ($\alpha = 0$), no spectral phase modulation occurs, yielding a pure spatial vortex aligned with the t-axis. As α increases initially, the increasing spatial chirp introduces stronger coupling between spatial and temporal coordinates, causing the vortex to rotate toward the *x*-axis. Simultaneously, spatial modulation (which arises from the spiral phase acting on the spatially dispersed spectrum) competes with spectral modulation (which arises from the same spiral phase acting on the frequency components). When spectral modulation becomes dominant at larger α, the orientation reverses and rotates back toward the *t*-axis, as quantified in **Fig. 4(a4)**.

**Figs. 4(b1-b4)** reveal a clear inverse relationship between focal length and tilt slope. As the objective focal length increases, the focal spot size enlarges while the pulse duration remains constant, causing the vortex orientation to approach the x-axis with the slope decreasing toward zero. The quantitative analysis in **Fig. 4(b4)** confirms that $k_t \propto 1/f$ (red line, **Fig. 4(b4)**), consistent with analytical prediction (Eq. (21) in **Methods**) as well as established pulse-front tilt scaling laws in Ref. [28].

**Figs. 4(c1-c4)** illustrate the more intricate dependence on input beam size ($W_x$). Changes in the input beam size not only modify the effective numerical aperture but also alter the spatial overlap between different spectral components, leading to a nonlinear relationship between beam size and STOV orientation. This is consistent with the analytical prediction (Eq. (21) in **Methods**), where $k_t$ depends of $W_x^2$.

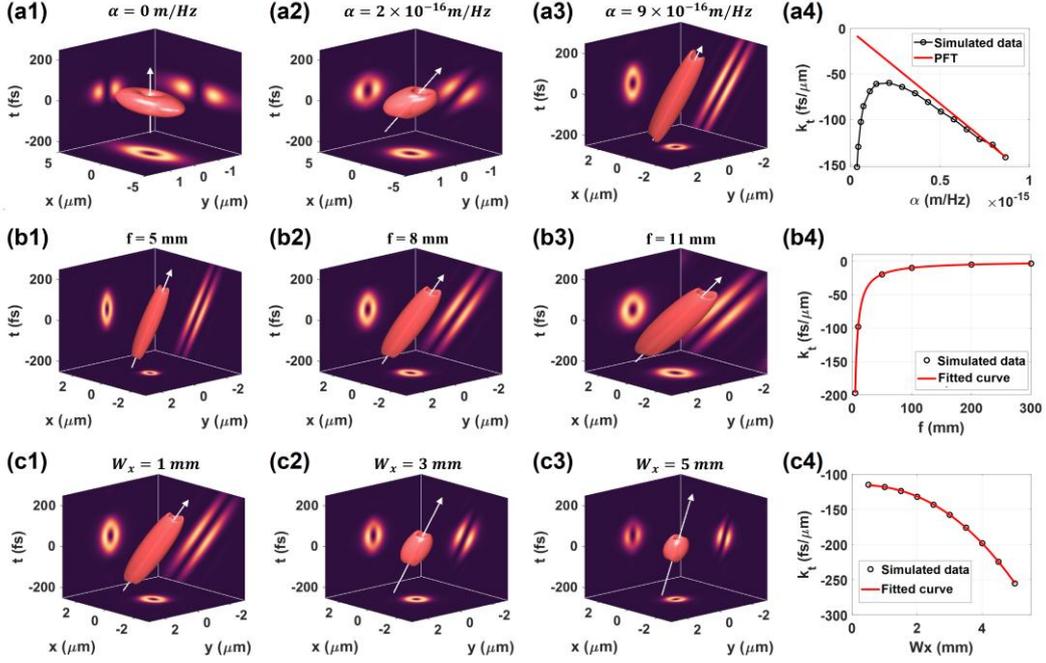

**Figure 4. Controlling the 3D OAM orientation of TF-STOV through system parameters.** (a) Tuning the spatial dispersion factor $\alpha$ adjusts the STOV orientation. Without spatial dispersion, a pure spatial vortex is generated. As the spatial dispersion increases, the spatiotemporal vortex component gradually emerges, tilting the overall angular momentum vector. The maximum dispersion is limited by the objective entrance pupil size. (b) Varying the objective focal length changes the focused spatial dimension, thereby rotating the STOV orientation in the spatiotemporal plane. Longer focal lengths produce smaller tilt angles approaching zero, increasing the relative contribution of the spatiotemporal vortex component. (c) The input beam size in the x-direction also modulates the vortex orientation by changing the spatial-to-temporal aspect ratio at focus. In all simulations, the default parameters are spatial dispersion factor $\alpha = 7 \times 10^{-16}\ m/Hz$, focal length $f = 8\ mm$, and input beam waist $W_x = 1\ mm$. Each row varies one parameter while the other two are held at their default values.

Collectively, these results demonstrate that our system enables continuous steering of the OAM vector from purely longitudinal ($k_t \to \infty$) to strongly transverse ($k_t \to 0$) configurations in 3D space-time domain. This tunability, achieved through straightforward adjustment of experimental parameters, highlights the versatility of our approach for generating and controlling three-dimensional OAM in focused geometries.

### 2.5 Experimental verification of TF-STOV under high NA conditions

Direct characterization of the spatiotemporal focal field is technically challenging, requiring sub-micron spatial sampling, femtosecond temporal resolution, and specialized interferometric diagnostics. Methods such as SEA TADPOLE with NSOM probes [29], TERMITES [30] for complete space–time reconstructions, or INSIGHT [31], are either complex or only suitable for very loosely focusing (i.e. NA < 0.1). Instead of using such methods, in this work, to further validate the robustness of spatiotemporally co-scaling TF-STOVs, we carried out femtosecond laser ablation experiments. Femtosecond ablation acts as a fluence-integrating probe, and the observed scanning electron microscopy (SEM) morphologies were benchmarked against predictions computed from the time-integrated intensity distribution.

The schematic of our experimental setup is plotted in Supplementary **Fig. S3**. The effective NA in this experiment is estimated to be 0.5. Ablation morphologies recorded on a silicon substrate at a series of axial positions (*z*-axis) were imaged by SEM and compared against numerical simulations, as shown in **Fig. 5**. For the STOV with topological charge $\ell = 1$ (**Fig. 5a**), at different defocus distance, the simulated 3D intensity distributions all show the characteristic tilted hollow structure with ring-shaped cross-sections in both the spatiotemporal (*y-t*) and spatial (*x-y*) planes. Due to the oblique orientation of the vortex, the time-integrated spatial intensity profiles exhibit a distinctive two-lobe pattern rather than a symmetric ring of the traditional spatial optical vortices. This arises from the temporal stacking of spatially displaced annular patterns as the tilted vortex sweeps across the material surface during the pulse duration. The experimental ablation patterns, characterized by SEM imaging at various defocus positions, show remarkable agreement with these time-integrated simulations, confirming successful generation of the tilted STOV.

In **Fig. 5a**, it is noted that the simulated time-integrated intensity exhibits a characteristic rotation of the two-lobe pattern along the propagation axis. This behavior originates from the Gouy phase dynamics of off-axis vortices. In our configuration, spatial dispersion displaces different frequency components transversely, causing each to carry a vortex at a different off-axis position. During focusing, each frequency component experiences a distinct Gouy phase shift due to the wavelength-dependent Rayleigh range, leading to differential phase accumulation, similar to the mechanism in Ref. [32-33]. Upon spectral recombination, these wavelength-dependent phase shifts manifest as a continuous rotation of the interference pattern with propagation distance. Such rotation serves as an additional experimental signature for identifying tilted STOVs under high NA condition.

We also compare the rotation in both simulations and experimental ablation patterns, as indicated by green dashed lines in **Fig. 5(a)**. At the focal plane ($z = 0$), both theory and experiment show no rotation. As the defocus distance increases, the lobe centers progressively rotate, and the experimental results agree quantitatively with theoretical predictions. This excellent correspondence demonstrates that the observed ablation patterns faithfully capture the spatiotemporal vortex structure.

As a first control experiment, we removed the spatial dispersion while retaining the vortex phase modulation ($\ell = 1$, **Fig. 5b**). In this configuration, the spiral phase is imprinted onto a conventional, non-spatially-dispersed beam before focusing, producing a pure spatial optical vortex rather than a STOV. The simulated three-dimensional intensity distributions confirm that the resulting donut-shaped structure is confined entirely to the *x-y* plane and exhibits no spatiotemporal coupling. Accordingly, the time-integrated intensity profiles display a well-defined annular distribution that remains rotationally symmetric across all defocus positions, in contrast to the two-lobe patterns observed in the STOV case (**Fig. 5a**). The SEM images of the ablation craters further corroborate this behavior, revealing circular ring-shaped morphologies consistent with conventional spatial vortices.

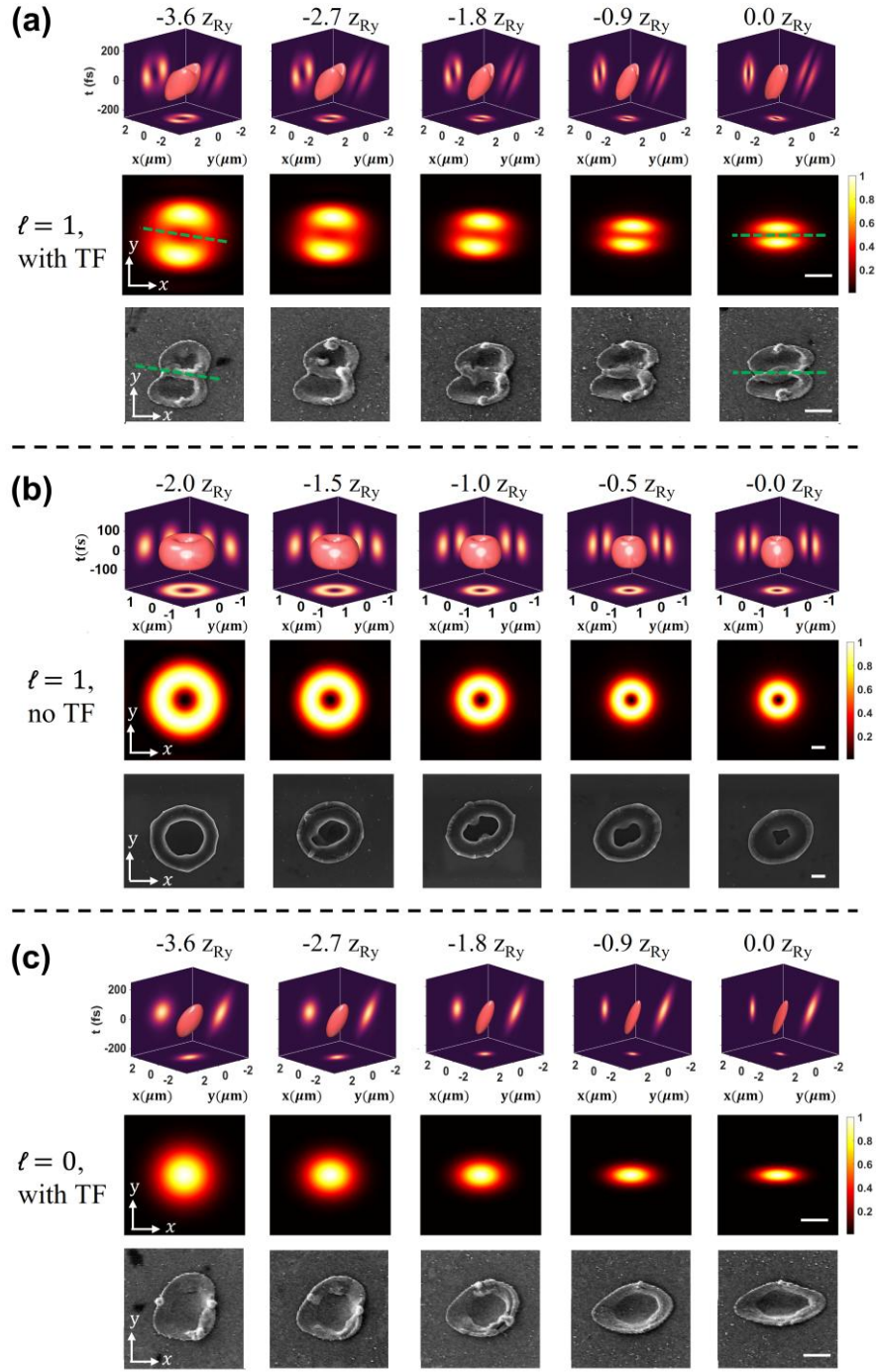

**Figure 5. Experimental validation of TF-STOV generation under high NA condition.** (a) First-order STOV ($\ell = 1$) generated with temporal focusing at different defocus distances. Top row: simulated 3D intensity distributions showing a tilted STOV with the characteristic ring structure in both the spatiotemporal plane (*y-t* plane) and the spatial plane (*x-y* plane). Middle row: time-integrated spatial intensity profiles, where the pulse-front tilt causes the ring to develop into two lobes due to temporal stacking of displaced annular patterns. Bottom row: SEM images of femtosecond laser ablation patterns, showing excellent agreement with the simulated time-integrated distributions. Scale bar: 1 μm. (b) Control experiment without temporal focusing ($\ell = 1$, no TF). In this case, the vortex phase is imprinted onto a non-spatially-dispersed beam before focusing, producing a pure spatial optical vortex. Top row: simulated 3D intensity distributions displaying a rotationally symmetric ring structure confined to the x-y plane with no spatiotemporal coupling. Middle row: time-integrated intensity profiles exhibiting a well-defined annular distribution. Bottom row: SEM images of ablation patterns confirming the circular ring morphology consistent with a conventional spatial vortex. Scale bar: 200 nm (c) Control experiment without spectral phase modulation ($\ell = 0$, with TF). Top row: simulated 3D intensity showing no ring structure in either spatiotemporal plane or spatial plane. Middle row: time-integrated profile exhibiting a near-Gaussian distribution due to the absence of a singularity. Bottom row: corresponding SEM characterization of ablation patterns, confirming the theoretical predictions. Scale bar: 1 μm.

As a second control experiment, we examined the case without spectral phase modulation ($\ell = 0$, **Fig. 5c**), which produces a zeroth-order field with no phase singularity. In the absence of spiral phase modulation, the intensity distribution resembles the pulse-front-tilted Gaussian beam reported by He et al. [28]. The simulated intensity distributions show no ring structure in either the spatiotemporal or spatial domains. Correspondingly, the time-integrated intensity exhibits a near-Gaussian distribution, and the resulting ablation patterns confirm this prediction with smooth, crater-like morphologies typical of conventional focused Gaussian beams. The ablation area for $\ell = 0$ exceeds that of $\ell = 1$, despite smaller focal spot sizes. This difference arises from the higher peak intensity at focus in the $\ell = 0$ case, where the absence of spectral phase modulation allows all frequency components to interfere constructively.

Together, the comparison among the three cases, namely STOV generation with both spatial dispersion and vortex phase ($\ell = 1$, with TF), pure spatial vortex without dispersion ($\ell = 1$, without TF), and no vortex phase ($\ell = 0$, with TF), demonstrates that the STOV is uniquely produced by the combination of spatial dispersion and spectral vortex phase modulation. The clear distinction in both simulation and experiment validates our ability to generate and control STOVs through spectral phase shaping in relatively high NA focusing configuration using our temporal focusing system.

### 2.6 Experimental verification of TF-STOV under low NA conditions

Since current full-field spatiotemporal measurement techniques are restricted to loosely focused geometries, we further validated the generation of TF-STOV under low-NA conditions by constructing an interferometric measurement system. The schematic of our experimental setup is plotted in Supplementary **Fig. S4**. The output of the femtosecond laser was split into a signal arm and a reference arm. In the signal arm, the beam first passed through a stretcher to acquire a controllable temporal chirp, and then entered a grating pair that introduced spatial dispersion along with a negative group-delay dispersion (GDD) (see **Methods**). By adjusting the stretcher, the positive chirp imposed on the signal beam was tuned to compensate the negative GDD accumulated through the grating pair, ensuring that the net GDD vanished at the focal plane and the pulse was recompressed to its transform-limited duration. After the grating pair, the spatially dispersed beam traversed a vortex retarder to imprint the spiral spectral phase, and was subsequently focused by a lens (f = 750 mm) onto a CCD camera (NA≈0.002). The reference arm provided a clean, unchirped beam free of spatial dispersion, which was combined with a small angle with the focused signal beam to produce spatial interference fringes recorded by the CCD. By scanning the relative time delay between the signal and reference pulses, a series of interferograms at different delays was acquired. A Fourier transform along the delay axis was then applied to isolate the positive first-order component, from which the complex electric-field amplitude of the signal beam was retrieved.

**Figure 6** presents the reconstructed results at four representative propagation distances from the focal plane. The three-dimensional spatiotemporal intensity profiles (**Fig. 6a**) reveal a donut-shaped structure at each propagation distance, confirming the successful generation of TF-STOV. Notably, due to the long focal length (f = 750 mm) employed in this low-NA configuration, the singularity slope $k_t$ is very small, which is estimated to be 2.7 fs/μm by fitting the singularity line in this experiment, consistent with the $k_t \propto 1/f$ scaling discussed in **Fig. 4(b)**. The y-t cross-sections (**Fig. 6b**) show that, as the defocus distance increases, the pulse duration τ(z) (indicated by white dashed lines) broadens progressively owing to incomplete temporal recompression. Importantly, the vortex phase singularity and the characteristic ring topology remain well preserved throughout. This behavior is further quantified in **Fig. 6c**, where the measured pulse duration τ(z) is plotted as a function of propagation distance z. The experimental data (black circles) are in good agreement with the theoretical fit (red curve), confirming the self-similar propagation behavior predicted by the TF-STOV scheme.

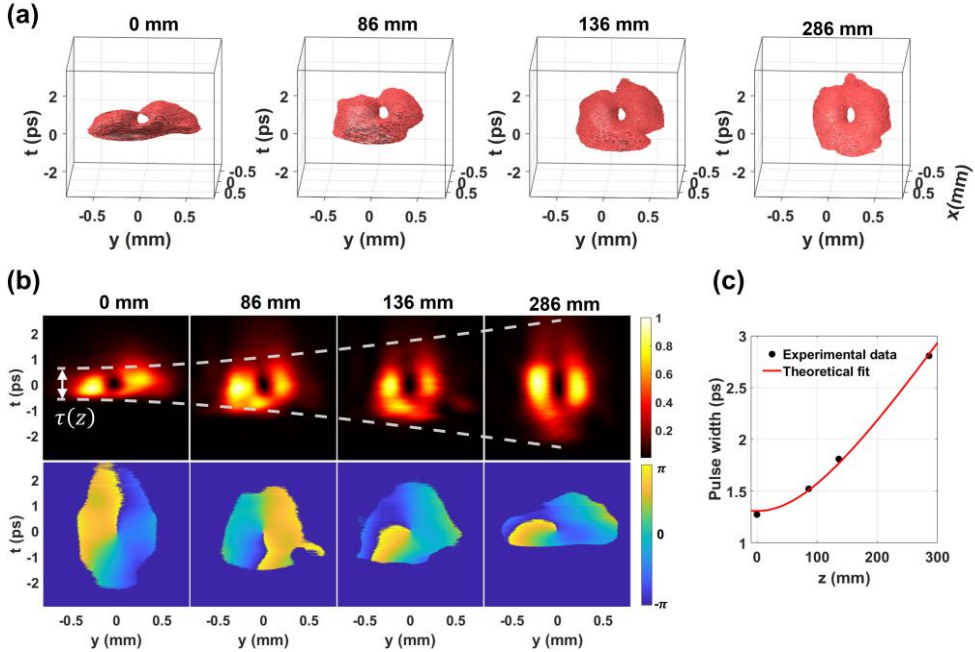

**Figure 6. Experimental validation of TF-STOV ($\ell = 1$) generation under low-NA conditions.** (a) Three-dimensional spatiotemporal intensity distributions reconstructed at four propagation distances (0, 86, 136, and 286 mm) from the focal plane. Due to the long focal length employed in this configuration, the singularity slope $k_t$ is very small. (b) Measured intensity (upper row) and phase (lower row) distributions in the y-t plane. As the defocus distance increases, the pulse duration τ(z) (indicated by white dashed lines) broadens progressively owing to incomplete temporal recompression, while the vortex phase singularity and the characteristic ring topology remain well preserved throughout propagation. (c) Quantitative analysis of the pulse duration (FWHM) as a function of propagation distance z. Black circles represent experimental data extracted from the y-t cross-sections in (b), and the red curve shows the theoretical fit $\tau(z) = \tau_0\sqrt{1 + z^2/z_{Rt}^2}$ with $\tau_0 = 1.3\ ps$ and $z_{Rt} = 149.5\ mm$.

## 3    Conclusion

In summary, we have demonstrated that the integration of spectral vortex phase modulation with temporal focusing provides a robust and tunable route to generating spatiotemporal optical vortices. By exploiting the intrinsic spatiotemporal coupling of temporal focusing, in which a grating pair forces the pulse to compress only at the geometric focus, the spatial and temporal dimensions evolve in lockstep throughout propagation. This eliminates the spatiotemporal astigmatism that troubles conventional 4f-based schemes, preserving the ring-shaped intensity profile and phase singularity in a self-similar form over an extended focal region. The angular dispersion inherent in temporal focusing further enables continuous control of the OAM orientation, smoothly rotating it from purely transverse to strongly longitudinal by adjusting the spatial dispersion, focal length, or input beam size.

These predictions are confirmed by two complementary experiments. Under high-NA focusing conditions, femtosecond laser ablation imprinting provides nanoscale-resolution spatial maps of the focal field, and systematic comparison with control cases without either spatial dispersion or vortex phase successfully identifies the STOV signature. Under low-NA conditions, interferometric field reconstruction directly retrieves the full three-dimensional complex amplitude. The reconstructed spatiotemporal profiles confirm that the long focal length yields a very small singularity slope $k_t$, consistent with the predicted $k_t \propto 1/f$ scaling. Moreover, the measured pulse duration τ(z) as a function of propagation distance is in good agreement with the theoretical prediction, and the vortex singularity and ring topology remain well preserved across a wide defocus range, corroborating the self-similar propagation behavior inherent to the TF-STOV scheme.

Looking forward, the integration of temporal focusing with structured spectral phase modulation extends well beyond STOV generation. Temporal focusing supplies a spatiotemporal coupling engine that binds spatial and temporal degrees of freedom, while spectral phase shaping provides a versatile topological and wavefront design toolkit. This synergy opens a pathway toward sculpting arbitrary spatiotemporal light fields, including higher-order vortices, multiplexed topological states, and spatiotemporal structures yet to be explored, all within a tightly focused, high-intensity geometry compatible with strong-field physics, ultrafast nonlinear optics, and nanoscale light–matter control.

# Method

### 1. Numerical simulation of TF-STOV generation using Fresnel diffraction theory

The field of a spatially dispersed pulse before the objective lens can be expressed as

$$A_1(x,y,\omega) = A_0 exp\left[-\frac{(\omega-\omega_0)^2}{W_\Omega^2}\right] exp\left\{-\frac{[x-\Delta x(\omega)]^2}{W_x^2} - \frac{y^2}{W_y^2}\right\}, \qquad (1)$$

where $A_0$ is a field amplitude, $\omega_0$ is the central frequency, $\sqrt{2ln2}W_\Omega$ is the full width at half maximum (FWHM) of the frequency spectrum of the pulse, $W_x$ and $W_y$ are the incident beam waists along x (spatial dispersing direction) and y directions, respectively, and $\Delta x(\omega) \approx \alpha(\omega - \omega_0)$ is the displacement of each spectral component with $\alpha$ being the spatial dispersion factor.

After a spiral phase is applied, the field can be written as

$$A_2(x,y,\omega) = A_1(x,y,\omega)exp(i\ell\varphi) \qquad (2)$$

where $\varphi = \tan^{-1}\left(\frac{y}{x}\right)$ is the azimuthal angle in the x-y plane, and $\ell$ is the topological charge.

The field after the objective can be written as

$$A_3(x,y,\omega) = A_2(x,y,\omega)exp\left(-ik\frac{x^2+y^2}{2f}\right) \qquad (3)$$

where $k = 2\pi c/\omega$ ($c$ is the light velocity in vacuum) and $f$ is the focal length of the objective.

Under the paraxial approximation, the laser field near the focus can be obtained by Fresnel diffraction:

$$A_4(x,y,z,\omega) = \frac{exp\,[ik(z+f)]}{i\lambda(z+f)} \iint_{-\infty}^{\infty} A_3(\xi,\eta,\omega) \, exp\left[ik\frac{(x-\xi)^2+(y-\eta)^2}{2(z+f)}\right] d\xi d\eta \qquad (4)$$

The intensity distribution in the time domain is obtained by the inverse Fourier transform of $A_4(x,y,\omega)$:

$$I(x,y,z,t) = |A_4(x,y,z,t)|^2 = \left|\int_{-\infty}^{\infty} A_4(x,y,z,\omega)exp(-i\omega t)d\omega\right|^2 \qquad (5)$$

Eqs. (1)-(5) are evaluated numerically without further approximation for all simulation results presented in this work.

### 2. Analytical solution of TF-STOV ($\ell = 1$)

For analytical deduction, we adopt two approximations: (1) Monochromatic approximation $k(\omega) \approx k_0 = \omega_0/c$. This is valid when the spectral width is small compared to the carrier frequency ($|\omega - \omega_0| \ll \omega_0$). This condition is well satisfied for femtosecond pulses, where typically $W_\Omega/\omega_0 \sim 10^{-2}$. (2) LG mode approximation: The spiral phase factor $exp(i\ell\varphi)$ in Eq. (2) introduces a $1/r$ singularity that prevents the Fresnel integral from being evaluated in close form. For $\ell = 1$, the spiral phase factor $e^{i\varphi} = (x+iy)/\sqrt{x^2+y^2} = (x+iy)/r$ is approximated by Laguerre-Gaussian mode,

$$e^{i\varphi} = \frac{x+iy}{r} \approx \frac{x+iy}{r_0} \qquad (6)$$

where $r_0$ is a normalization constant absorbed into $A_0$ hereafter. The two forms agree closely within the beam, and the discrepancy at large $r$ is exponentially suppressed by the Gaussian envelope. This representation has been widely adopted in the analytical treatment of spatiotemporal optical vortices.

After these approximations, Eq. (2) becomes (for $\ell = 1$):

$$A_2(x,y,\Omega) = A_0(x+iy)exp\left[-\frac{\Omega^2}{W_\Omega^2}\right] exp\left\{-\frac{[x-\alpha\Omega]^2}{W_x^2} - \frac{y^2}{W_y^2}\right\} \qquad (7)$$

where $\Omega = \omega - \omega_0$.

Defining $R = f + z$ as the propagation distance and $\frac{1}{Q} = \frac{1}{f} - \frac{1}{R}$ as the defocus parameter, Eq. (4) becomes:

$$A_4(x,y,z,\Omega) = \frac{k_0 \exp[ik_0 R]}{2\pi i R} \iint_{-\infty}^{\infty} A_2(\xi,\eta,\Omega) exp\left(-ik\frac{\xi^2+\eta^2}{2f}\right) exp\left[ik\frac{(x-\xi)^2+(y-\eta)^2}{2R}\right] d\xi d\eta$$

$$= \frac{k_0 \exp[ik_0 R]}{2\pi i R} \cdot \exp\left[\frac{ik_0(x^2+y^2)}{2R}\right] \cdot \mathcal{L}(x,y,z,\Omega) \tag{8}$$

where $\mathcal{L}$ is the double integral.

$$\mathcal{L}(x,y,z,\Omega) = \iint_{-\infty}^{\infty} A_2(\xi,\eta,\omega) \cdot \exp\left[-\frac{ik_0(\xi^2+\eta^2)}{2Q}\right] \cdot \exp\left[-\frac{ik_0}{R}(x\xi+y\eta)\right] d\xi d\eta \tag{9}$$

where $\frac{1}{Q} = \frac{1}{f} - \frac{1}{R}$ is the defocus parameter.

We further introduce a centered coordinate $u = \xi - \alpha\Omega$ and apply two standard Gaussian integral:

$$\int_{-\infty}^{\infty} e^{-Pu^2+su} du = \sqrt{\frac{\pi}{P}} e^{s^2/(4P)}, \quad \int_{-\infty}^{\infty} u e^{-Pu^2+su} du = \frac{s}{2P}\sqrt{\frac{\pi}{P}} e^{s^2/(4P)} \tag{10}$$

Eq. (9) becomes

$$\mathcal{L}(x,y,z,\Omega) = \exp\left(-i\frac{k_0 x\alpha\Omega}{R}\right) \exp\left(-\frac{\Omega^2}{W_\Omega^2}\right) \exp\left(-i\frac{k_0\alpha^2\Omega^2}{2Q}\right) \frac{\pi}{\sqrt{P_x P_y}} \exp\left(\frac{s_x^2}{4P_x} + \frac{s_y^2}{4P_y}\right)\left(\frac{s_x}{2P_x} + i\frac{s_y}{2P_y} + \alpha\Omega\right) \tag{11}$$

with

$$\begin{cases} P_x = \frac{1}{W_x^2} + i\frac{k_0}{2Q} \\ P_y = \frac{1}{W_y^2} + i\frac{k_0}{2Q} \\ s_x = -i\left(\frac{k_0 x}{R} + \frac{k_0\alpha\Omega}{Q}\right) \\ s_y = -i\frac{k_0 y}{R} \end{cases} \tag{12}$$

Substituting Eq. (12) into Eq. (9) generates

$$A_4(x,y,z,\Omega) = \frac{k_0 \exp[ik_0 R]}{2\pi i R} \exp\left[\frac{ik_0(x^2+y^2)}{2R}\right] \exp\left(-i\frac{k_0 x\alpha\Omega}{R}\right) \exp\left(-\frac{\Omega^2}{W_\Omega^2}\right) \exp\left(-i\frac{k_0\alpha^2\Omega^2}{2Q}\right) \cdot$$

$$\frac{\pi}{\sqrt{P_x P_y}} \exp\left(\frac{s_x^2}{4P_x} + \frac{s_y^2}{4P_y}\right)\left(\frac{s_x}{2P_x} + i\frac{s_y}{2P_y} + \alpha\Omega\right) \tag{12}$$

Now we can separate the frequency-domain field in Eq. (12):

$$A_4(x,y,z,\Omega) = \mathcal{C}(x,y,z) \exp\left(-\frac{\Omega^2}{W_{\text{eff}}^2}\right) \exp(-i\phi_1 \Omega)[\mathcal{B}_0(x,y,z) + \mathcal{B}_1 \Omega] \tag{13}$$

with

$$\begin{cases} \frac{1}{W_{\text{eff}}^2} = \frac{1}{W_\Omega^2} + i\frac{k_0\alpha^2}{2Q} + \frac{k_0^2\alpha^2}{4Q^2 P_x} \\ \mathcal{C}(x,y,z) = \frac{k_0 \exp[ik_0 R]}{2\pi i R} \exp\left[\frac{ik_0(x^2+y^2)}{2R}\right] \frac{\pi}{\sqrt{P_x P_y}} \exp\left(-\frac{k_0^2 x^2}{4P_x R^2}\right) \exp\left(-\frac{k_0^2 y^2}{4P_y R^2}\right) \\ \mathcal{B}_0(x,y,z) = -i\frac{k_0}{2R}\left(\frac{x}{P_x} - i\frac{y}{P_y}\right) \\ \mathcal{B}_1 = \left(1 - i\frac{k_0}{2P_x Q}\right)\alpha \\ \phi_1 = \frac{k_0\alpha x}{R}\left(1 + i\frac{k_0}{2P_x Q}\right) \end{cases}$$

Eq. (13) reveals that the focal field is a product of a Gaussian envelope and a vortex factor. The vortex factor contains two distinct terms that carry clear physical meaning. The first term, $\mathcal{B}_0(x,y,z)$, depends only on the spatial coordinates and represents a purely spatial vortex component. The second term, $\mathcal{B}_1 \Omega$, is proportional to the frequency offset $\Omega$. Upon Fourier transformation to the time domain (Eq. (14) as below) constituting a spatiotemporal vortex component that carries transverse OAM. The coexistence of these terms means that the focal field naturally decomposes into co-propagating spatial and spatiotemporal vortex components, whose relative weight is governed by the system parameters such as $\alpha$, $f$ and $W_x$.

We can also see that the real part of $1/W_{\text{eff}}^2$ controls the effective bandwidth. At focus ($1/Q = 0$), it equals $1/W_\Omega^2$. When away from focus, the bandwidth narrows. That's the origin of pulse broadening temporally.

To performing the $\Omega \to t$ Fourier transform of Eq. (13), we use two integrals:

$$\begin{cases} \int_{-\infty}^{\infty} e^{-ax^2+bx} dx = \sqrt{\frac{\pi}{a}} e^{b^2/(4a)} \\ \int_{-\infty}^{\infty} x e^{-ax^2+bx} dx = \frac{b}{2a} \int_{-\infty}^{\infty} e^{-ax^2+bx} dx \end{cases}$$

The time-domain field amplitude becomes:

$$A_4(x,y,z,t) = \int_{-\infty}^{\infty} A_4(x,y,z,\Omega) e^{-i\Omega t} d\Omega$$

$$= \mathcal{C}(x,y,z) W_{\text{eff}} \sqrt{\pi} \exp\left(-\frac{W_{\text{eff}}^2 T^2}{4}\right) \left[\mathcal{B}_0(x,y,z) - \frac{W_{\text{eff}}^2 T \alpha}{2}\left(i + \frac{k_0}{2QP_x}\right)\right] \quad (14)$$

with $T = \phi_1 + t$.

The intensity distribution can be obtained:

$$I(x,y,z,t) = |A_4(x,y,z,t)|^2$$

$$= |\mathcal{C}(x,y,z)|^2 \pi |W_{\text{eff}}|^2 \exp\left(-\frac{\text{Re}(W_{\text{eff}}^2) T^2}{2}\right) \left|\mathcal{B}_0(x,y,z) - \frac{W_{\text{eff}}^2 T \alpha}{2}\left(i + \frac{k_0}{2QP_x}\right)\right|^2 \quad (15)$$

3. **Analytical interpretation of temporal focusing**

From the Gaussian envelop $\exp\left(-\frac{\text{Re}(W_{\text{eff}}^2) T^2}{2}\right)$ in Eq. (15), the pulse duration (1/e field half-duration) evolution can be obtained:

$$\tau(z) = 2/\sqrt{\text{Re}(W_{\text{eff}}^2)} \quad (16)$$

Under near-focus condition,

$$Q \approx f^2/z \quad (17)$$

$$\tau(z) \approx \tau_0 \sqrt{1 + z^2/z_{Rt}^2} \quad (18)$$

with $\tau_0 = 2/W_\Omega$ and $z_{Rt} = \frac{f^2 \tau_0}{k_0 \alpha W_x}$ as the temporal Rayleigh range.

Self-similar propagation requires equal temporal and y-direction spatial Rayleigh range $z_{Rt} = z_{Ry}$, which gives:

$$\alpha W_\Omega = \frac{W_y^2}{W_x} \quad (19)$$

When satisfied, TF-STOV maintains optimal topological stability during propagation.

4. **Analytical interpretation of OAM orientation control**

At the focal plane ($z = 0$), the vortex term in Eq. (14) becomes,

$$V = \left[\mathcal{B}_0(x,y,z) - \frac{W_{\text{eff}}^2 T \alpha}{2}\left(i + \frac{k_0}{2QP_x}\right)\right] = [by - i(ax + gT)] \quad (20)$$

with

$$\begin{cases} b = -\dfrac{k_0 W_y^2}{2f} \\ a = \dfrac{k_0 W_x^2}{2f} \\ g = \dfrac{W_\Omega^2 \alpha}{2} \end{cases}$$

The field vanished where both real and imaginary parts of the vortex term are zero:

$$by = 0 \text{ and } ax + gT = 0$$

This defines a line in $(x, y, T)$-space:

$$y = 0 \text{ and } T = -\frac{a}{g}x$$

The vortex core is a straight line tilted in the $x - T$ plane.

At the focal plane, $T = \phi_1 + t = \frac{k_0 \alpha x}{f} + t$. If we define the singularity slope $k_t = \frac{dt}{dx}$, we have

$$k_t = -\frac{1}{f}\left[\frac{k_0 W_x^2}{W_\Omega^2 \alpha} + k_0 \alpha\right] \tag{21}$$

Eq. (21) explains the three experimental dependencies:

(1) $\alpha$ dependence: Competition between two terms, $k_t$ minimum at $\alpha_{opt} = W_x/W_\Omega$;
(2) $f$ dependence: Linear $k_t \propto 1/f$ relationship
(3) $W_x$ dependence: Nonlinear monotonic increase through $W_x^2$ term.

## 5. Calculation of intensity correlation coefficients

To assess propagation self-similarity, we compute intensity correlations as follows. Each spatiotemporal distribution $I(y, t, z)$ is normalized to [0, 1] and spatially rescaled by τ(z)/τ(0) to account for expected diffractive scaling, where τ(z) is the pulse duration at position z. The 2D Pearson correlation coefficient between the rescaled distribution and the reference at $z = 0$ quantifies structural preservation. The correlation coefficient is defined as [17]:

$$r = \frac{\iint I(y,t,z) \cdot I(y,t,z=0) dy dt}{\iint I(y,t,z) dy dt \cdot \iint I(y,t,z=0) dy dt} \tag{6}$$

## 6. Femtosecond laser ablation experiment

The experimental setup is illustrated in Fig. S3. A chirped pulse amplification (CPA) system delivers laser pulses with a center wavelength of 800 nm, spectral bandwidth of 10 nm (FWHM), and linear polarization along the *x*-direction. To pre-compensate for dispersion introduced by downstream optical elements (neutral density filters, objective lens, and gratings), the pulse is positively pre-chirped by adjusting the internal compressor in the CPA system. Optimal dispersion compensation is achieved when air ionization at the objective focus produces maximum plasma luminescence, corresponding to minimum pulse duration (~100 fs). The input beam size in both *x*- and *y*-directions is controlled by a two-dimensional adjustable slit. The incident pulse energy is adjusted using neutral density (ND) filters.

The beam is directed through a parallel grating pair (1200 lines/mm) with a separation distance of 210 mm at an incident angle of 45° to introduce spatial dispersion along the *x*-direction. The spatially dispersed spectrum is subsequently reflected by a phase-only spatial light modulator (SLM, Hamamatsu X15213-02) that applies a spiral phase mask exp(iℓφ) with topological charge ℓ. After phase modulation, a 100× objective lens (nominal NA = 0.9, Olympus) recombines all frequency components at the focal plane. Since the beam does not overfill the objective's back aperture, the effective numerical aperture is reduced to $NA_{eff} \approx 0.4$, which is used in the simulations presented in Fig. 5.

Silicon substrate is mounted on a motorized three-axis translation stage for ablation pattern acquisition at different axial positions z. The ablation morphology is subsequently characterized using a scanning electron microscope (Zeiss, GeminiSEM 560).

For comparison with ablation experiments, we computed the time-integrated intensity distribution from Eq. (5):

$$I_{int}(x,y,z) = \int I(x,y,z,t)dt \tag{7}$$

This quantity represents the total fluence deposited on the material surface, which determines the ablation morphology for femtosecond pulses in the low-fluence regime.

7. Interferometric measurement experiment

The experimental setup is illustrated in Fig. S4. A Ti:sapphire laser amplifier (Legend Elite HE, Coherent) delivers femtosecond pulses with a central wavelength of 800 nm and a pulse duration of 120 fs. The output is split by a beam splitter (BS1) into two arms. In the signal arm, the beam first passes through a Martinez-type pulse stretcher to acquire a positive temporal chirp, and then enters a single-pass grating compressor that introduces both the required spatial dispersion and a negative group-delay dispersion (GDD). The Martinez-type pulse stretcher is carefully designed to compensate this unwanted negative GDD. After the compressor, a vortex retarder imprints the spiral spectral phase onto the spatially dispersed beam, which is subsequently focused by a lens (L3) to generate the TF-STOV. In the reference arm, the beam propagates without spatial dispersion or temporal chirp and is combined at a small angle with the focused TF-STOV to produce interference fringes on a CCD camera. The relative time delay between the two arms is precisely controlled by a motorized translation stage (V-408, PI). By recording interferograms at a series of delay values and applying a Fourier transform along the delay axis to isolate the positive first-order component, the full spatiotemporal complex-field amplitude of the TF-STOV is retrieved.

The Martinez-type pulse stretcher employs a double-pass configuration consisting of two reflective gratings (G1 and G2; 1200 lines/mm, blazed at 750 nm) and a 4f telescope (lenses L1 and L2, each with a focal length of 200 mm). Both gratings are oriented with an incident angle of 40°. The lens separation is 400 mm and the grating-to-lens distance is approximately 86.5 mm. The effective distance of the pulse stretcher z was estimated to be ~454 mm. The positive GDD introduced by the stretcher can be expressed as

$$GDD = \frac{\lambda^3 z}{2\pi c^2 a^2 (\cos\theta_d)^2} \tag{8}$$

where $\lambda$ is the central wavelength, $z$ is the effective distance of the stretcher, $a$ is the grating constant, and $\theta_d$ is the first-order diffraction angle.

The temporally chirped beam exiting the stretcher is then directed into a single-pass grating compressor composed of two reflective gratings (G3 and G4; 1200 lines/mm, blazed at 800 nm) at normal incidence, separated by a perpendicular distance of 10.5 mm. The compressor serves a dual role: it generates the spatial dispersion necessary for temporal focusing while simultaneously introducing a negative GDD:

$$GDD = -\frac{\lambda^3 d}{2\pi c^2 a^2 (\cos\theta_d)^3} \tag{9}$$

where $d$ is the perpendicular distance of the gratings.

The GDD contributed by the compressor is estimated to be $-6.24 \times 10^5$ fs², which is compensated by the positive GDD from the Martinez-type stretcher, ensuring that the net GDD vanishes at the focal plane and the pulse is recompressed to its transform-limited duration.

**Acknowledgements**

This research was supported by the National Natural Science Foundation of China (62575184, 62375177), Shenzhen Science and Technology Program (JCYJ20210324120403011, RCJC20210609103232046, RCJC20200714114435063, JCYJ20241202124428038, JCYJ20250604181103005), Shenzhen Medical Research Fund (D250403001), Research Team Cultivation Program of Shenzhen University (2023QNT014), and Shenzhen University 2035 Initiative (2023B004), Natural Science Foundation of Guangdong Province (2026A1515010428). The authors acknowledge the Photonics Center of Shenzhen University for the technical support.

**Data availability**

The data underlying the results presented in this paper are not publicly available at this time but may be obtained from the authors upon reasonable request.

**Conflict of interests**

The authors declare no conflicts of interest.